\documentclass[reprint,aps,prb,amsmath,superscriptaddress,bibnotes,longbibliography]{revtex4-2}

\usepackage{mathptmx,newtxtext,newtxmath,xspace}
\usepackage{amsbsy,bm,bbold}
\usepackage{graphicx,color,xcolor}
\usepackage{fancyhdr}
\usepackage[colorlinks=true, 
linkcolor=blue, 
urlcolor=blue,
citecolor=blue]{hyperref}
\usepackage{psfrag}
\usepackage{enumitem}
\newcommand{\iu}{{i\mkern1mu}}
\usepackage{soul}
\begin{document}
	
	
	\title{Effect of random antiferromagnetic exchange on the spin waves in a three-dimensional Heisenberg ferromagnet}    
	
	
	\widetext
	\date{\today}
	
\author{S. Hameed}
\thanks{Present address: Max Planck Institute for Solid State Research, Heisenbergstraße 1, 70569 Stuttgart, Germany}
\affiliation{School of Physics and Astronomy, University of Minnesota, Minneapolis, MN 55455, USA}

\author{Z. Wang}
\thanks{Present address: Center for Correlated Matter and School of Physics, Zhejiang University, Hangzhou 310058, China}
\affiliation{School of Physics and Astronomy, University of Minnesota, Minneapolis, MN 55455, USA}

\author{D. M. Gautreau}
\affiliation{School of Physics and Astronomy, University of Minnesota, Minneapolis, MN 55455, USA}
\affiliation{Chemical Engineering and Materials Science, University of Minnesota, Minneapolis, MN 55455, USA}

\author{J. Joe}
\affiliation{School of Physics and Astronomy, University of Minnesota, Minneapolis, MN 55455, USA}
\author{K. P. Olson}
\thanks{Present address: Department of Materials Science and Engineering, Northwestern University, Evanston, IL 60201, USA;}
\affiliation{Chemical Engineering and Materials Science, University of Minnesota, Minneapolis, MN 55455, USA}
\author{S. Chi}
\affiliation{Neutron Scattering Division, Oak Ridge National Laboratory, Oak Ridge, Tennessee 37831, USA}
\author{P. M. Gehring}
\affiliation{NIST Center for Neutron Research, National Institute of Standards and Technology, Gaithersburg, Maryland 20899, USA}
\author{T. Hong}
\affiliation{Neutron Scattering Division, Oak Ridge National Laboratory, Oak Ridge, Tennessee 37831, USA}
\author{D. M. Pajerowski}
\affiliation{Neutron Scattering Division, Oak Ridge National Laboratory, Oak Ridge, Tennessee 37831, USA}
\author{T. J. Williams}
\affiliation{Neutron Scattering Division, Oak Ridge National Laboratory, Oak Ridge, Tennessee 37831, USA}
\author{Z. Xu}
\affiliation{NIST Center for Neutron Research, National Institute of Standards and Technology, Gaithersburg, Maryland 20899, USA}
\author{M. Matsuda}
\affiliation{Neutron Scattering Division, Oak Ridge National Laboratory, Oak Ridge, Tennessee 37831, USA}
\author{T. Birol}
\affiliation{Chemical Engineering and Materials Science, University of Minnesota, Minneapolis, MN 55455, USA}
\author{R. M. Fernandes}
\affiliation{School of Physics and Astronomy, University of Minnesota, Minneapolis, MN 55455, USA}
\author{M. Greven}
\affiliation{School of Physics and Astronomy, University of Minnesota, Minneapolis, MN 55455, USA}

	\begin{abstract}
	Neutron scattering is used to study spin waves in the three-dimensional Heisenberg ferromagnet YTiO$_3$, with spin-spin exchange disorder introduced $via$ La-substitution at the Y site. No significant changes are observed in the spin-wave dispersion up to a La concentration of 20\%. However, a strong broadening of the spectrum is found, indicative of shortened spin-wave lifetimes. Density-functional theory calculations predict minimal changes in exchange constants as a result of average structural changes due to La substitution, in agreement with the data. The absence of significant changes in the spin-wave dispersion, the considerable lifetime effect, and the reduced ordered magnetic moment previously observed in the La-substituted system are qualitatively captured by an isotropic, nearest-neighbor, three-dimensional Heisenberg ferromagnet model with random antiferromagnetic exchange. We therefore establish Y$_{1-x}$La$_x$TiO$_3$ as a model system to study the effect of antiferromagnetic spin-exchange disorder in a three-dimensional Heisenberg ferromagnet.
		
	\end{abstract}
	\maketitle
	
	Understanding how disorder affects magnetic ground states is the subject of considerable research, not only due to fundamental scientific interest \cite{KORENBLIT1988109,Elliot1974}, but also because of its potential to advance technologies such as spintronics~\cite{Macdonald2005} and magnonics~\cite{Barman2021}. Arguably the most powerful probe of magnetism is neutron scattering, which enables comprehensive measurements of magnetic order and excitations, and therefore provides pivotal information regarding the spin Hamiltonian that governs the system \cite{Shirane2002}. 
	
	The effects of non-magnetic site dilution in magnetic materials has been well-studied theoretically and experimentally ($e.g.$,~\cite{Cowley1972,Elliott1974,Wagner1978,Nakayama1994,Vajk2002,Rafael2011}). A prominent example is site dilution in two-dimensional Heisenberg antiferromagnets~\cite{Vajk2002}. Other examples include studies of dilute ferromagnetic (FM) and antiferromagentic (AFM) transition metal fluorides~\cite{Cowley1972,Wagner1978,Nakayama1994}. Spin-exchange disorder in the form of intermixed AFM and FM interactions has also been studied theoretically, in the context of spin-glasses~\cite{Binder1986}. The relevant models typically consider a Gaussian distribution of spin-exchange, spanning both FM and AFM interactions, but centered around a FM exchange~\cite{Binder1986}. Such models allow only for FM and spin-glass ground states. The case of a binary distribution of FM and AFM interactions, which allows for both FM and AFM ground states, has also been explored theoretically to some extent~\cite{Feigelman1979,Medvedev1978,Ginzburg1979}. However, to the best of our knowledge, there exist no measurements of spin waves in experimental realizations of such a spin system with a binary distribution of exchange constants.

	The rare-earth titanates RETiO$_3$ (RE being a rare-earth ion) provide a unique platform with which to study the effects of such a binary spin-exchange distribution in a three-dimensional (3D) system. These perovskites are Mott insulators, with a spin-$\frac{1}{2}$ 3$d^1$ electronic configuration. They feature a GdFeO$_3$-type distorted structure, with a Ti-O-Ti bond angle that is less than 180$^{\text{o}}$. The nature of the magnetic ground state depends on the RE ionic radius, and hence the Ti-O-Ti bond angle. Materials with smaller RE ionic radii (\textit{i.e.}, smaller Ti-O-Ti bond angles) exhibit FM ground states ($e.g.$, Y, Gd, Dy), whereas materials with larger RE ionic radii (\textit{i.e.}, larger Ti-O-Ti bond angles) exhibit AFM ground states ($e.g.$, La, Sm, Nd)~\cite{Mochizuki2004}. Mixing different RE ions therefore enables the exploration of the effects of admixing FM and AFM spin exchanges in a 3D system \cite{Zhentao2022}. Spin-wave spectra in FM YTiO$_3$ and AFM LaTiO$_3$ were measured previously and shown to be in good agreement with a nearly isotropic 3D Heisenberg FM and AFM model, respectively~\cite{Ulrich2002,Keimer2000}. Furthermore, unlike most other RETiO$_3$ systems, Y$^{3+}$ and La$^{3+}$ lack a magnetic moment, which offers a distinct advantage: the system's magnetic properties stem solely from the Ti$^{3+}$ ions.
	
	\begin{figure}
	\includegraphics[width=0.35\textwidth]{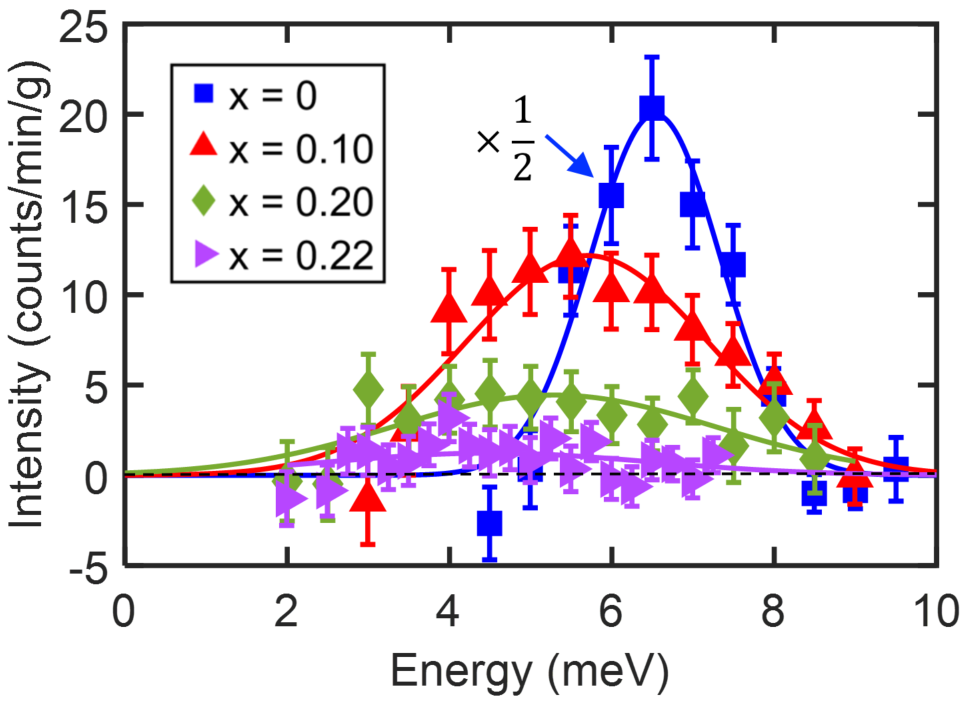}
		\caption{Representative energy scans at $(0,0,1)_{\text{o}}$ (with high-temperature data subtracted) for $x = 0$, 0.10 and 0.20, along with respective gaussian fits. The data for $x = 0$ are scaled down by a factor of 2. See~\cite{SM} for additional energy and momentum scans.}
		\label{fig:sw001}
	\end{figure}

	\begin{figure*}
		\includegraphics[width=\textwidth]{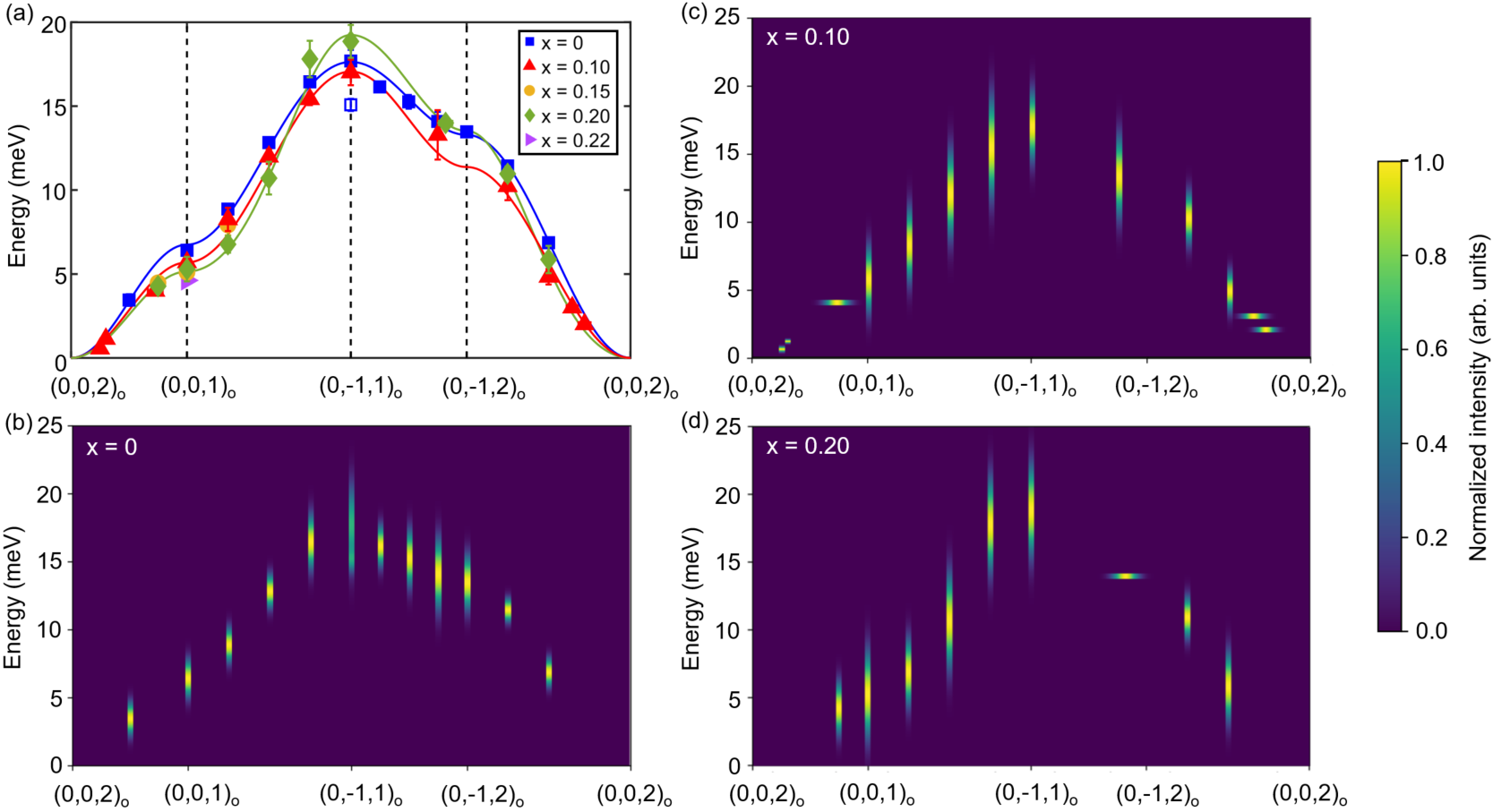}
		\caption{(a) Spin-wave peak energies at low temperature. For $x = 0$, two peaks were resolved at $(0,-1,1)_{\text{o}}$, likely as a result of a coupling with an optical phonon branch~\cite{Ulrich2002}. The open blue square indicates the second peak. The lines are guides to the eye. (b,c,d) Spin wave intensities at low temperature for (b) $x = 0$, (c) $x = 0.10$, and (d) $x = 0.20$, obtained from gaussian fits to energy scans, as described in the text. The data elongated along the horizontal direction were obtained from momentum scans. For each scan, the peak intensity is normalized to 1. The low-energy data at $(0,0,1.7)_{\text{o}}$ and $(0,0,1.75)_{\text{o}}$ for $x = 0.10$ have a small width in energy as these were taken with a higher energy resolution at a cold-neutron triple-axis spectrometer. For $x = 0$, the intensities of the two peaks resolved at $(0,-1,1)_{\text{o}}$ were normalized such that the sum of the peak intensities is 1. The data for $x = 0$ and $x = 0.20$ and the data at $(0,0,1.7)_{\text{o}}$ and $(0,0,1.75)_{\text{o}}$ for $x = 0.10$ were obtained at 1.5 K. The remaining data were obtained at 4 K.}
		\label{fig:spinwaves}
	\end{figure*}

	In this Letter, we study spin waves in the Mott-insulating spin-$\frac{1}{2}$ 3D Heisenberg FM YTiO$_3$, with spin-exchange disorder induced $via$ substitution of La at the Y site. We present data up to a La substitution level of $x = 0.22$. We find minimal changes in the spin-wave dispersion, whereas a strong broadening of the spectrum is observed with increasing La substitution. Complementary density functional theory (DFT) calculations reveal that the average structural changes induced by moderate La substitution are expected to cause only a modest decrease in spin-exchange parameters, in agreement with the data. Spin-wave calculations based on a nearest-neighbor, isotropic, Heisenberg FM model with added random AFM spin-exchange reveal a shortened spin-wave lifetime, a reduced FM ordered moment, and subtle changes in the spin-wave dispersion with increasing disorder, in good qualitative agreement with the experimental results. Y$_{1-x}$La$_{x}$TiO$_{3}$ is therefore a model system for the study of AFM spin-exchange disorder effects in a 3D Heisenberg ferromagnet.


	\begin{figure}
		\includegraphics[width=0.3\textwidth]{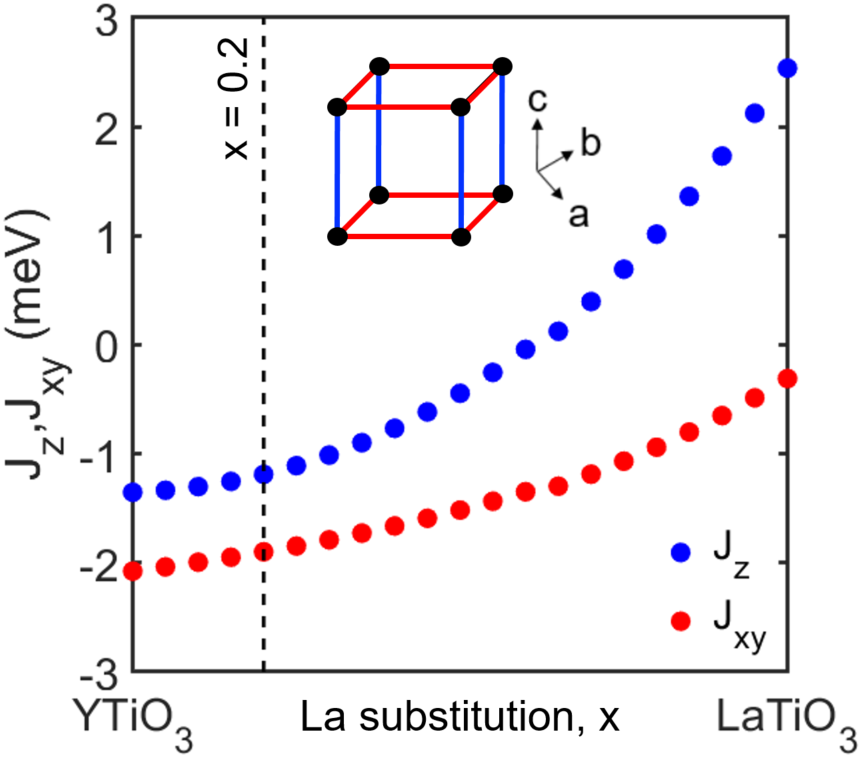}
		\caption{Effective spin-spin exchange parameters as a function of La-substitution determined from DFT. Each La concentration is simulated by interpolating between the known crystal structures for YTiO$_3$ and LaTiO$_3$. $J_{xy}$ and $J_z$ represent spin-exchange parameters between nearest-neighbor Ti ions in the $ab$-plane and along the $c$-axis, respectively (see inset).}
		\label{fig:DFT}
	\end{figure}
	
	\begin{figure*}
		\includegraphics[width=0.9\textwidth]{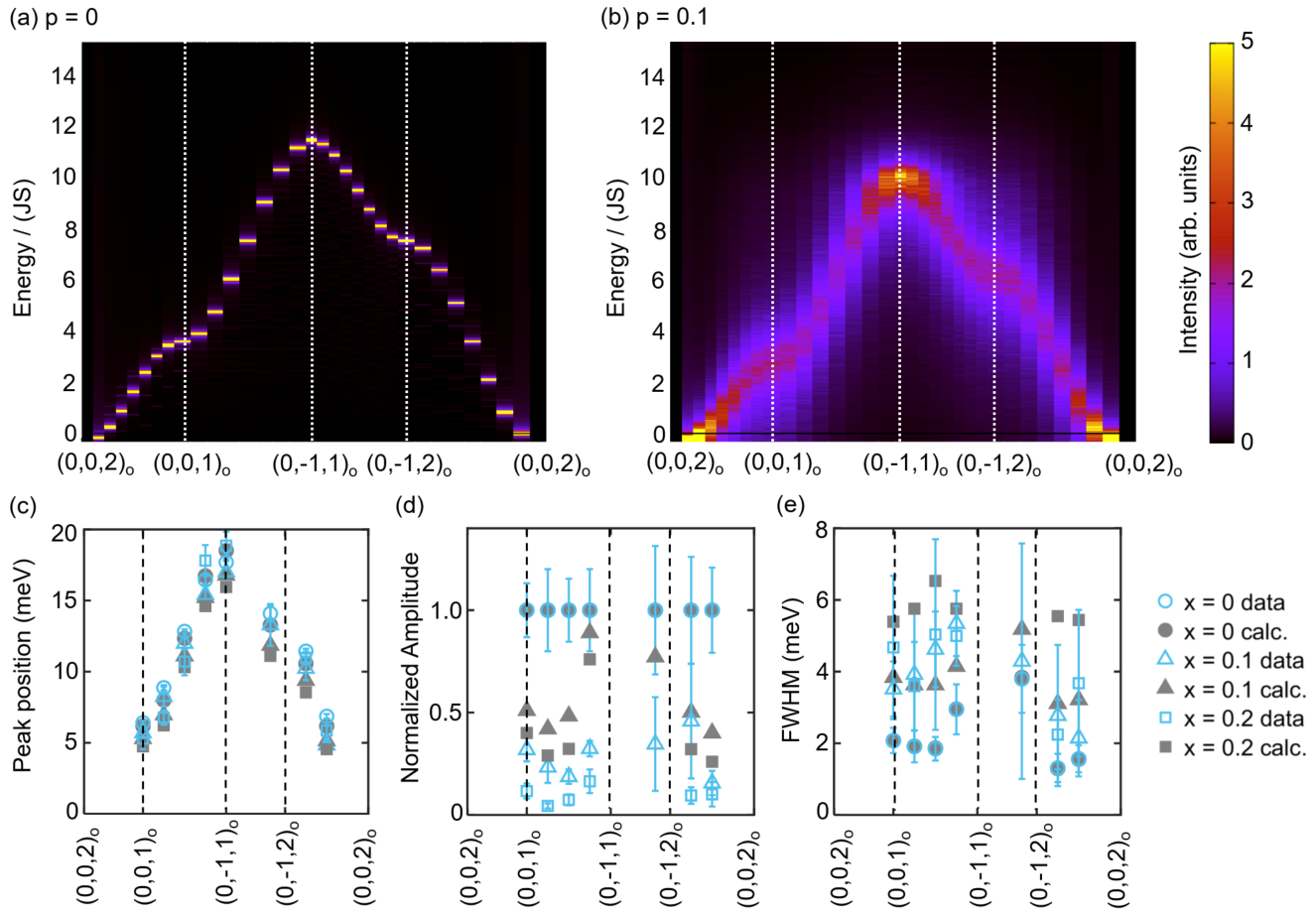}
		\caption{(a) Calculated dynamic spin structure factor $\mathcal{S}(\bm{k},\omega)$ of model~\eqref{eq:4} for (a) $p = 0$ and (b) $p = 0.1$, obtained with $L=16$ and temperature $T = 0.1J$. Here, $S = \frac{1}{2}$. (c,d,e) Comparison of the (c) peak positions, (d) amplitude, and (e) full-width-at-half-maximum (FWHM) of the calculated and measured spin-wave results at different wave vectors. We set $p = x/2$ for the comparison. The calculated spin-wave spectrum is scaled to best fit (for $p = 0$) the experimental results for YTiO$_3$. At each wave vector, the amplitude is normalized to the $x = p = 0$ value. The theoretical results are convoluted with the instrument resolution to determine the FWHM. Only results at wave vectors for which data are available for multiple La substitution levels are displayed in (c-e). The comparison at (0,-1,1)$_{\text{o}}$ is omitted in (d,e) since a coupling to phonons complicates the experimental spin-wave response \cite{Ulrich2002}. For the comparison at (0,-1,1)$_{\text{o}}$ in (c), we use the higher-energy spin-wave peak obtained for $x = 0$.}
		\label{fig:disorder}
	\end{figure*}
	
	Single crystals of Y$_{1-x}$La$_{x}$TiO$_{3}$ were melt-grown with the optical floating-zone technique and well-characterized post-growth, as described elsewhere~\cite{HameedGrowth2021}. Triple-axis neutron spectroscopy was performed with the HB-1 and HB-3 thermal-neutron triple-axis spectrometers and the CG-4C cold-neutron triple axis spectrometer at the High Flux Isotope Reactor, located at Oak Ridge National Laboratory, and with the BT-7 thermal-neutron triple axis spectrometer, located at the NIST Center for Neutron Research. We studied $x= 0$ and $x = 0.10$ single-crystals with masses in the 3-4 g range, and an $x = 0.20$ sample composed of 20 co-aligned crystals with a total mass of $\sim$ 12 g. The larger sample mass for $x = 0.20$ was needed to detect the weaker signal due to the reduced ordered magnetic moment and strongly damped spin waves. Limited data were also collected on $x= 0.15$ and $x = 0.22$ single crystals, with masses of $\sim$ 1.5 g each, and an $\sim$ 1.5 g $x = 0.30$ sample that consists of 11 co-aligned crystals. The fixed final energies of the scattered neutrons were 14.7 and 30.5 meV for the thermal-neutron triple-axis measurements and 4.5 meV for the cold-neutron triple-axis measurements. We used collimations of $48'-80'- $sample$ - 80'- 120'$ at HB-1 and HB-3, open$ -80' - $sample$ - 80' - 120'$ at BT-7, and open$ -$open$ -$sample$ - 80' - $open at CG-4C. PG filters were used at HB-1, HB-3 and BT-7 and a cooled Be filter was used at CG-4C to eliminate higher-order neutrons. A $^4$He-cryostat and a closed-cycle-refrigerator were used to reach temperatures down to 1.5 K and 4 K, respectively. The samples were mounted in the $(0KL)$ scattering plane, and the spin-wave excitations were characterized using energy and momentum scans. Each scan was performed twice: at a temperature well below the magnetic transition temperatures (1.5 - 4 K), and at a temperature well above the magnetic transition temperatures (15 - 40 K).
	

	Figure~\ref{fig:sw001} shows representative energy scans measured at $(001)_{\text{o}}$ for $x = 0, 0.10$ and 0.20. Here, the subscript “o” refers to the orthorhombic reciprocal space notation, which we use throughout this paper. The data obtained in the paramagnetic state were subtracted from the low-temperature data in order to better isolate the magnetic contribution to the inelastic signal in the magnetically-ordered state. Clear spin-wave signals are observed for all samples. A strong decrease in intensity is observed with increasing La substitution.
	
	At nonzero La concentrations, the signal-to-background ratio was not high enough to accurately extract intrinsic spin-wave lifetimes from a deconvolution with the instrument resolution (see supplementary material~\cite{SM}). Therefore, fits of the net (high-temperature data subtracted) data to a simple Gaussian profile were carried out in all cases. Examples of resultant fits are shown as solid lines in Fig.~\ref{fig:sw001} and in~\cite{SM}. The peak positions thus obtained are used to determine the spin-wave dispersion, as shown in Fig.~\ref{fig:spinwaves}(a). The spin-wave spectrum for YTiO$_3$ is in good agreement with prior work~\cite{Ulrich2002}. 
	
	For YTiO$_3$, we found the energy scans to be nearly resolution-limited, indicative of very long spin-wave lifetimes. The data shown in Fig.~\ref{fig:spinwaves}(b)-(d) for $x = 0$, 0.10, and 0.20 are plotted such that the intensity maximum in each case is set to one. We clearly observe that the spin waves broaden in energy with increasing La substitution, a result that is indicative of shortened lifetimes. 
	
	To shed light on the behavior of the spin-wave spectrum with increasing $x$, we perform DFT calculations that assume that the main effect of La substitution is to induce structural changes in Y$_{1-x}$La$_x$TiO$_3$. Thus, we approximate the crystal structure of Y$_{1-x}$La$_x$TiO$_3$ by interpolating between the YTiO$_3$ and LaTiO$_3$ structures. In particular, for a given value of $x$, the lattice parameters and internal coordinates of Y$_{1-x}$La$_x$TiO$_3$ are obtained by linearly interpolating those for bulk YTiO$_3$ and LaTiO$_3$ crystal structures. Taking 21 equally-spaced values of $x$ between $x=0$ and $x=1$, we obtain a series of interpolated structures. In each structure, we calculate the energy of 16 collinear spin configurations from first principles (see \cite{SM} for details). These energies are then fit to a nearest-neighbor Heisenberg model to obtain the nearest-neighbor exchange parameters $J_{xy}$ and $J_z$ (see inset to Fig.~\ref{fig:DFT}) as a function of La concentration, $x$. In all calculations, Y is used as the rare-earth ion. This assumption is valid if the effect of La substitution is largely steric in nature. The results are presented in Fig. \ref{fig:DFT}, where we show how $J_{xy}$ and $J_z$ vary with La concentration. Clearly, up to $x = 0.20$, only a modest decrease ($\sim$ 10\%) in $|J_{xy}|$ and $|J_z|$ is expected, consistent with the minimal changes observed in the spin-wave energy scale with increasing La substitution.
	
\begin{figure}
	\includegraphics[width=0.35\textwidth]{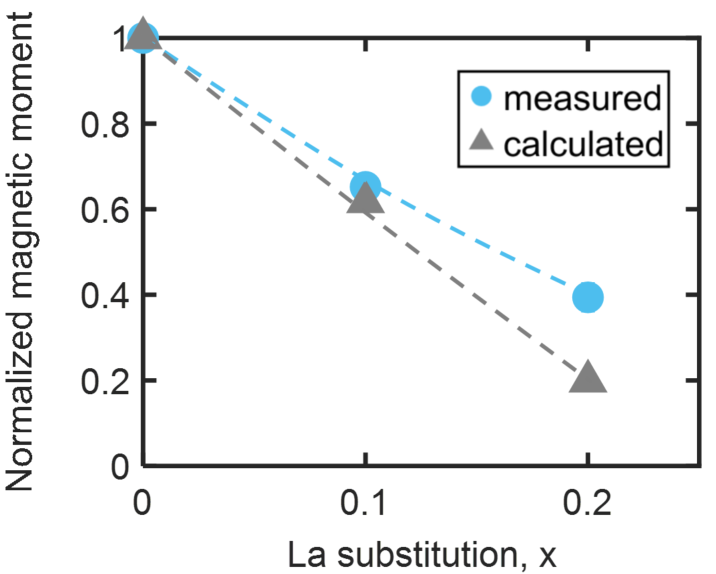}
	\caption{Calculated FM moment $\sqrt{\left[ \langle m^2 \rangle \right]_\text{av}}$ obtained for the disordered Heisenberg model with $L=16$ and $T/J=0.1$ (the errorbars are smaller than the symbol size), compared with prior data for Y$_{1-x}$La$_x$TiO$_3$ (from \cite{HameedYLa2021}). We set $p = x/2$ for the comparison. The FM moment values are normalized to the value at $x = p = 0$.}
	\label{fig:disorder2}
	\end{figure}

	In these DFT calculations, we considered only the average structural changes induced by La substitution. Such changes are not expected to drive the shortened spin-wave lifetimes that we observe in our data. In order to understand this, the effect of disorder resulting from the presence of atoms of differing sizes needs to be considered. Since LaTiO$_3$ is an antiferromagnet, in a simple picture, admixing La at the Y site in FM YTiO$_3$ induces AFM spin-exchange into the FM host. For simplicity, we consider an isotropic nearest-neighbor Heisenberg ferromagnet~\cite{Ulrich2002} with AFM Heisenberg interactions randomly admixed into it. Specifically, we consider the classical Heisenberg model on an $L \times L \times L$ simple cubic lattice,
	
	\begin{equation}
	\mathcal{H} = \sum_{\langle i,j \rangle}J_{ij}\bm{S}_i \cdot \bm{S}_j,
	\label{eq:4}
\end{equation}
where $J_{ij}$ is the nearest-neighbor magnetic exchange drawn from a binary distribution: $J_{ij} = -J$ (FM) with probability $1-p$, and $J_{ij} = 5J$ (AFM) with probability $p$. The ratio of the FM and AFM exchange constants is chosen to be $1:5$, to match the ratio of the Heisenberg spin exchanges in YTiO$_3$ ($\sim 3$ meV) \cite{Ulrich2002} and LaTiO$_3$ ($\sim 15.5$ meV) \cite{Keimer2000}. 
The dynamic spin structure factor was calculated on a $L=16$ cubic lattice using replica exchange Monte Carlo with heat-bath and over-relaxation updates~\cite{HukushimaK1996a,MiyatakeY1986,CreutzM1987,BrownFR1987}, followed by the equation-of-motion method~\cite{AlbenR1975}, and averaged over 50 disorder realizations (see \cite{SM} for details). The calculated spin-wave spectra are displayed in Fig.~\ref{fig:disorder}(a,b) for $p = 0$ and $p = 1$, respectively. A strong broadening in the spin-wave intensity is observed for nonzero $p$, while the zone-boundary energy remains approximately constant. This is in qualitative agreement with our data. Experimentally, we cannot rule out the possibility that the La ions cluster together. If such clustering were present, one would expect that the local Ti-O-Ti bond angle closely resembles that of LaTiO$_3$ and hence lead to an AFM spin exchange between the neighboring spin-$\frac{1}{2}$ Ti$^{3+}$ ions. We expect one AFM spin exchange bond for every two La ions in this picture. Motivated by this feature, and to make the comparison with experiment more quantitative, we relate $p$ and $x$ according to $p \approx x/2$. Despite the simplicity of the model, it qualitatively captures the key experimental observations with increasing $x$: (i) minimal changes in the spin-wave dispersion (Fig.~\ref{fig:disorder}(c)); (ii) a strong decrease of the spin-wave amplitude (Fig.~\ref{fig:disorder}(d)); (iii) a strong reduction in the spin-wave lifetime (Fig.~\ref{fig:disorder}(e)); and (iv) a strong decrease in the FM ordered moment with increasing $x$ (Fig.~\ref{fig:disorder2}). Although the experimental trends are reasonably captured by our model, there are quantitative deviations. For instance, the experimental spin-wave amplitude falls off more strongly than the calculations (Fig.~\ref{fig:disorder}(d)). The deviation near the zone boundary is particularly severe, most likely due to magnon-phonon crossing \cite{Ulrich2002}, which is not considered in our model. A more detailed theoretical description is beyond the scope of this work in light of the experimental constraint of low signal-noise ratios even with a large sample mass. Nevertheless, given the good qualitative agreement of the calculations with the data, we can conclude that the magnetic properties of Y$_{1-x}$La$_x$TiO$_3$ in the La substitution range up to $x \sim 0.2$ are dominated by effective AFM spin-exchange disorder introduced by the La substitution. In prior theoretical work on FM systems with admixed AFM interactions, it was suggested that local AFM spin waves form \cite{Ginzburg1979}. Attempts to find such local modes experimentally at the zone boundary were unsuccessful \cite{SM}. However, we cannot rule out the possibility that such spin waves are too weak in amplitude to be observable in our measurements.
	
	To summarize, we used neutron scattering to measure spin waves in the 3D Heisenberg ferromagnet Y$_{1-x}$La$_x$TiO$_3$. No significant changes are observed in the spin-wave dispersion. In contrast, a strong broadening and weakening of the spin-wave intensity is observed with increasing La concentration. We find good qualitative agreement with calculations for an isotropic nearest-neighbor FM Heisenberg model with randomly admixed AFM spin-exchange. We therefore conclude that, at substitution levels up to $x \approx 0.2$, Y$_{1-x}$La$_x$TiO$_3$ is a model 3D Heisenberg ferromagnet with AFM spin-exchange disorder.
	
	The work at University of Minnesota was funded by the Department of Energy through the University of Minnesota Center for Quantum Materials, under DE-SC0016371. Parts of this work were carried out in the Characterization Facility, University of Minnesota, which receives partial support from the NSF through the MRSEC (Award Number DMR-2011401) and the NNCI (Award Number ECCS-2025124) programs. A portion of this research used resources at the High Flux Isotope Reactor, a DOE Office of Science User Facility operated by the Oak Ridge National Laboratory. We acknowledge the support of the National Institute of Standards and Technology, U.S. Department of Commerce, in providing the neutron research facilities used in this work.

	
	\bibliography{refs}
	
	\widetext
	\clearpage
	
	\begin{center}
		\textbf{\large Supplemental Material}
	\end{center}

Here we document raw spin-wave data at all measured La concentrations (Section~\ref{rawspinwave}), raw energy scans showing the absence of AFM spin waves in $x = 0.30$ (Section~\ref{afmswsearch}), a comparison between YTiO$_3$ spin-wave spectra obtained via instrument resolution deconvolution and simple gaussian fits (Section~\ref{ytodeconv}), details regarding the DFT calculations (Section~\ref{dftcalc}) and Monte Carlo simulations (Section~\ref{montecarlo}).

	\clearpage
	\setcounter{equation}{0}
	\setcounter{figure}{0}
	\setcounter{table}{0}
	\setcounter{page}{1}
	\makeatletter
	\renewcommand{\theequation}{S\arabic{equation}}
	\renewcommand{\thetable}{S\arabic{equation}}
	\renewcommand{\thefigure}{S\arabic{figure}}
	\renewcommand{\thesection}{S\arabic{section}}
	\renewcommand{\bibnumfmt}[1]{[S#1]}
	\section{Raw spin-wave data}
	\label{rawspinwave}
	
	In this Section, we present raw energy and momentum scans used to determine spin-wave dispersions presented in the main text. In most cases, the difference between a low-temperature scan and a high-temperature scan is fit to a simple gaussian profile. In these cases, two separate plots are presented, one with the raw low-temperature (red filled circles) and high-temperature (blue filled circles) scans, and one with the difference data (black filled circles) fitted to a gaussian profile (solid black line). In a few cases, the background is clearly temperature dependent, and therefore the low-temperature data were directly fit to a gaussian profile (also shown by solid black lines). 
	
	\subsection{$x = 0$}
	
	\begin{figure}[!htb]
		\includegraphics[width=0.6\textwidth]{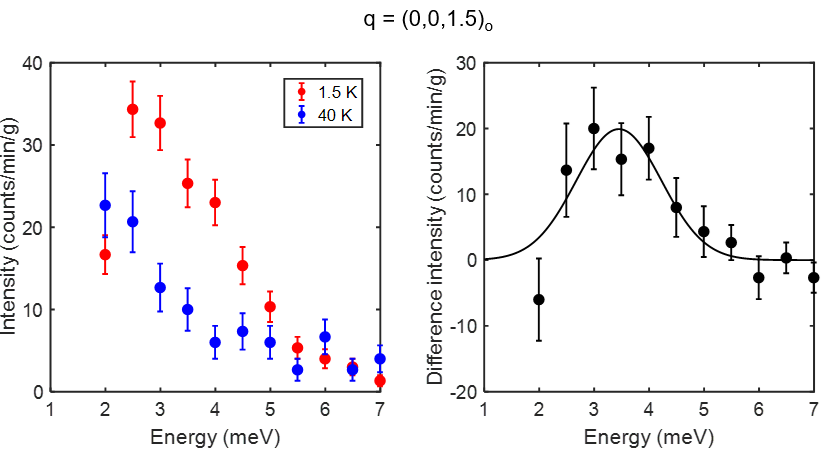}
	\caption{Left: Raw energy scans for $x = 0$ and $(0,0,1.5)_\text{o}$ at 1.5 K (red circles) and 40 K (blue circles). Right: Difference between the raw scans (black circles) and fit to gaussian profile (solid black line).}
		\label{fig:YTO_001P5}
	\end{figure}
	
	\begin{figure}[!htb]
		\includegraphics[width=0.3\textwidth]{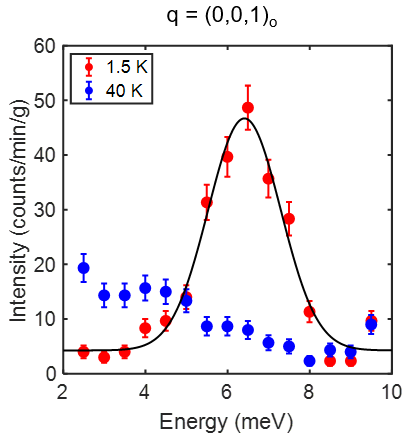}
	\caption{Left: Raw energy scans for $x = 0$ and $(0,0,1)_\text{o}$ at 1.5 K (red circles) and 40 K (blue circles). Right: Difference between the raw scans (black circles) and fit to gaussian profile (solid black line).}
		\label{fig:YTO_001}
	\end{figure}
	
	\begin{figure}[!htb]
	\includegraphics[width=0.6\textwidth]{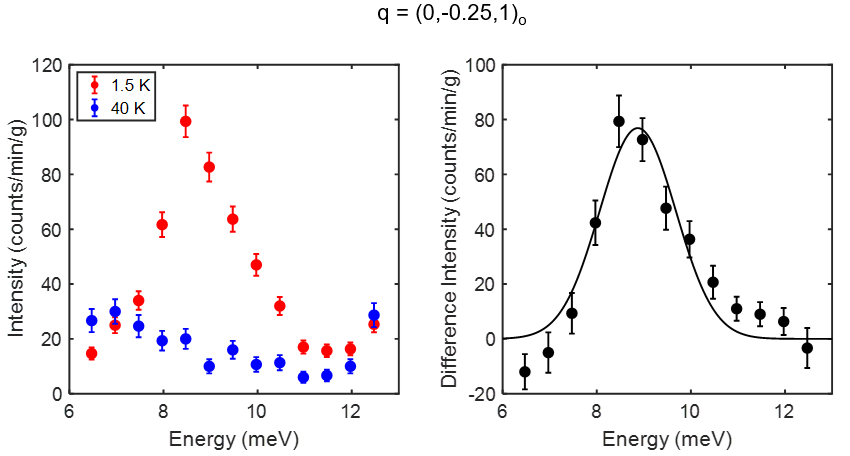}
	\caption{Left: Raw energy scans for $x = 0$ and $(0,-0.25,1)_\text{o}$ at 1.5 K (red circles) and 40 K (blue circles). Right: Difference between the raw scans (black circles) and fit to gaussian profile (solid black line).}
	\label{fig:YTO_00P251}
	\end{figure}

\begin{figure}[!htb]
	\includegraphics[width=0.6\textwidth]{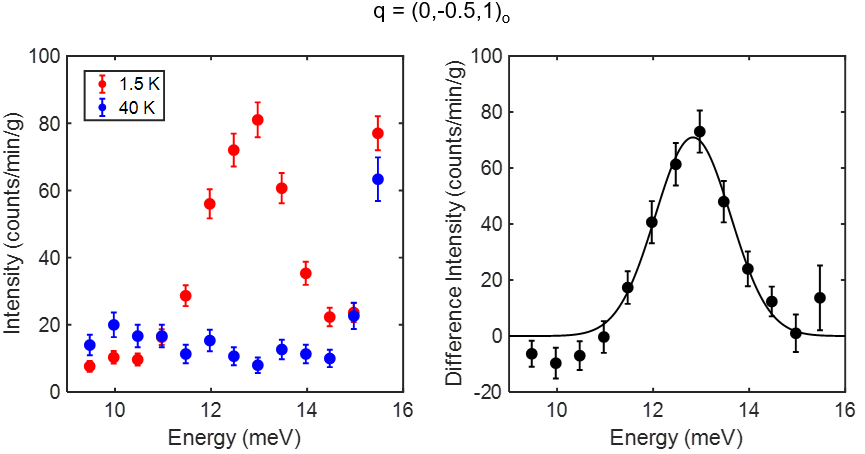}
	\caption{Left: Raw energy scans for $x = 0$ and $(0,-0.5,1)_\text{o}$ at 1.5 K (red circles) and 40 K (blue circles). Right: Difference between the raw scans (black circles) and fit to gaussian profile (solid black line).}
	\label{fig:YTO_00P51}
\end{figure}

\begin{figure}[!htb]
	\includegraphics[width=0.6\textwidth]{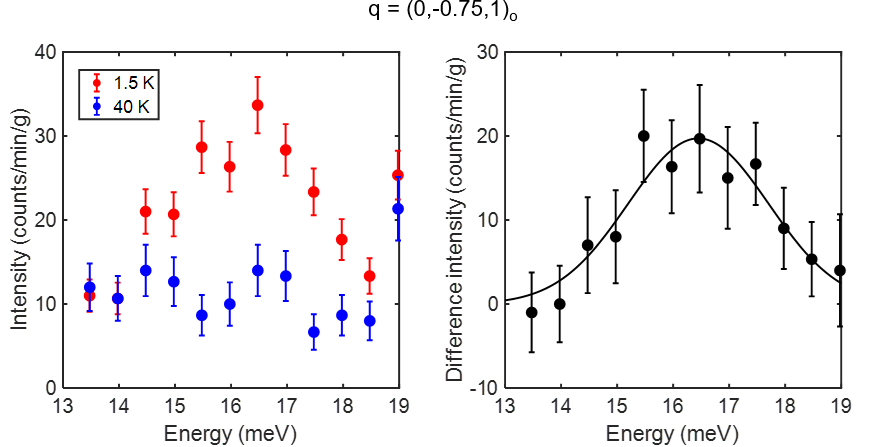}
	\caption{Left: Raw energy scans for $x = 0$ and $(0,-0.75,1)_\text{o}$ at 1.5 K (red circles) and 40 K (blue circles). Right: Difference between the raw scans (black circles) and fit to gaussian profile (solid black line).}
	\label{fig:YTO_00P751}
\end{figure}

\begin{figure}[!htb]
	\includegraphics[width=0.3\textwidth]{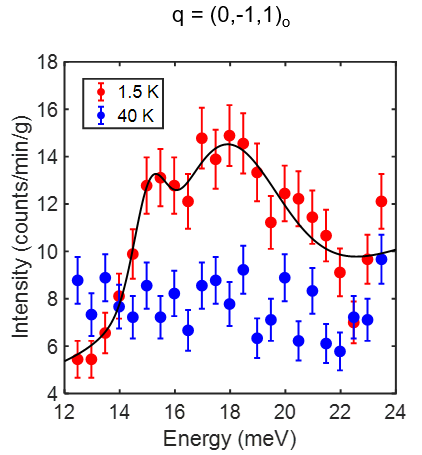}
	\caption{Left: Raw energy scans for $x = 0$ and $(0,-1,1)_\text{o}$ at 1.5 K (red circles) and 40 K (blue circles). Right: Difference between the raw scans (black circles) and fit to gaussian profile (solid black line).}
	\label{fig:YTO_011}
\end{figure}

\begin{figure}[!htb]
	\includegraphics[width=0.6\textwidth]{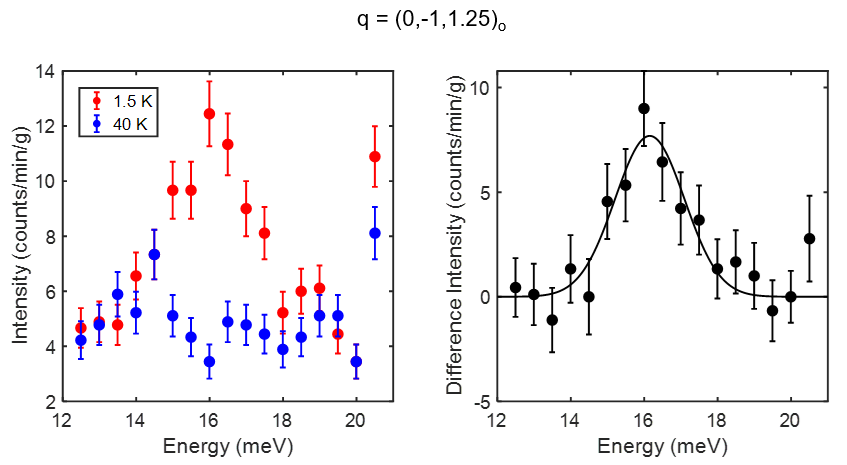}
	\caption{Left: Raw energy scans for $x = 0$ and $(0,-1,1.25)_\text{o}$ at 1.5 K (red circles) and 40 K (blue circles). Right: Difference between the raw scans (black circles) and fit to gaussian profile (solid black line).}
	\label{fig:YTO_011P25}
\end{figure}

\begin{figure}[!htb]
	\includegraphics[width=0.6\textwidth]{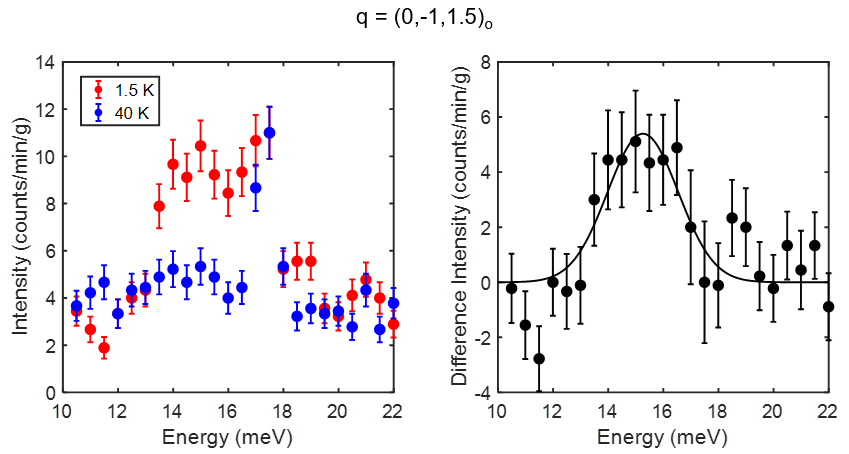}
	\caption{Left: Raw energy scans for $x = 0$ and $(0,-1,1.5)_\text{o}$ at 1.5 K (red circles) and 40 K (blue circles). Right: Difference between the raw scans (black circles) and fit to gaussian profile (solid black line).}
	\label{fig:YTO_011P5}
\end{figure}

\begin{figure}[!htb]
	\includegraphics[width=0.6\textwidth]{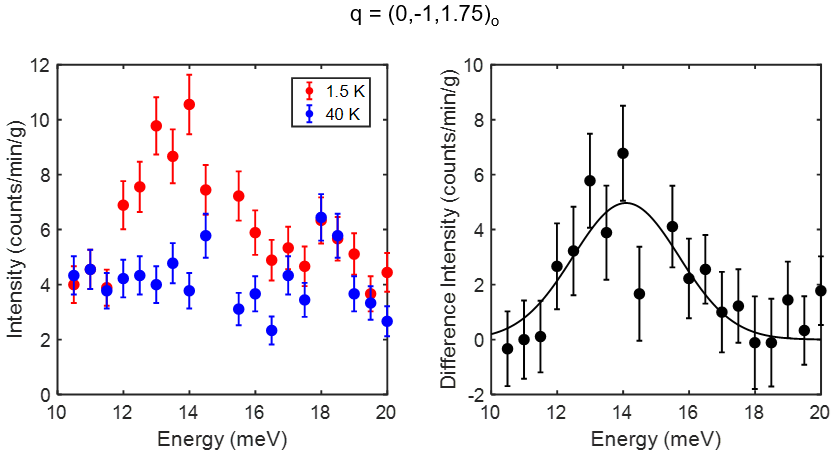}
	\caption{Left: Raw energy scans for $x = 0$ and $(0,-1,1.75)_\text{o}$ at 1.5 K (red circles) and 40 K (blue circles). Right: Difference between the raw scans (black circles) and fit to gaussian profile (solid black line).}
	\label{fig:YTO_011P75}
\end{figure}

\begin{figure}[!htb]
	\includegraphics[width=0.6\textwidth]{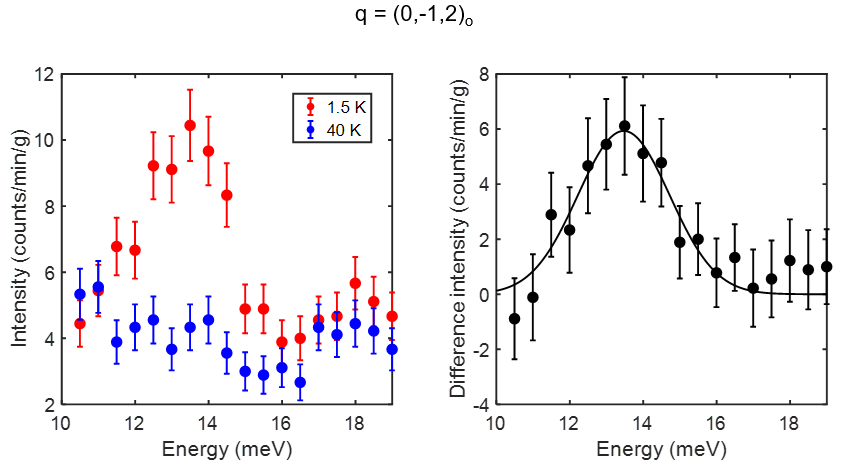}
	\caption{Left: Raw energy scans for $x = 0$ and $(0,-1,2)_\text{o}$ at 1.5 K (red circles) and 40 K (blue circles). Right: Difference between the raw scans (black circles) and fit to gaussian profile (solid black line).}
	\label{fig:YTO_012}
\end{figure}

\begin{figure}[!htb]
	\includegraphics[width=0.6\textwidth]{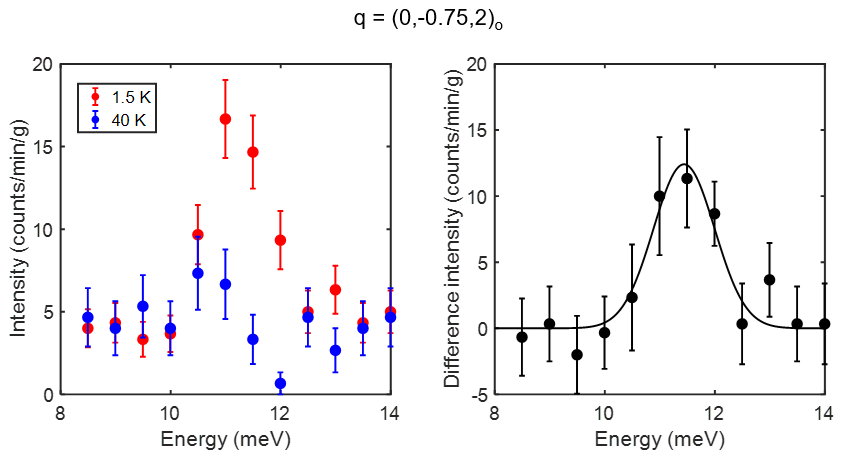}
	\caption{Left: Raw energy scans for $x = 0$ and $(0,-0.75,2)_\text{o}$ at 1.5 K (red circles) and 40 K (blue circles). Right: Difference between the raw scans (black circles) and fit to gaussian profile (solid black line).}
	\label{fig:YTO_00P752}
	\end{figure}

	\begin{figure}[!htb]
	\includegraphics[width=0.6\textwidth]{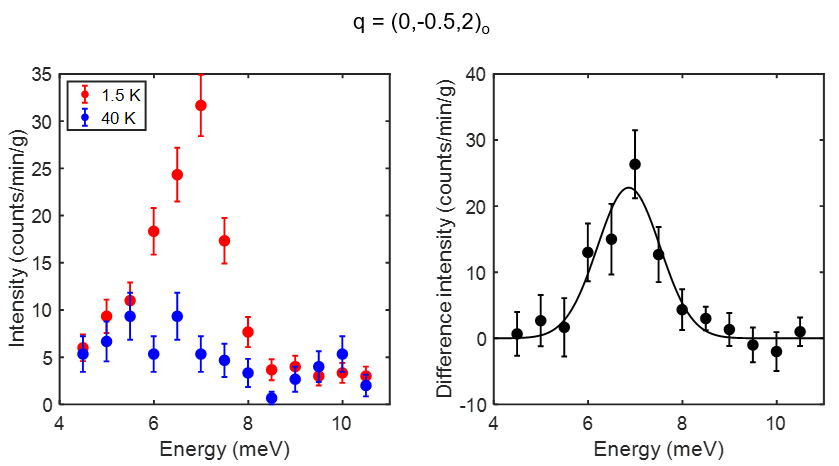}
	\caption{Left: Raw energy scans for $x = 0$ and $(0,-0.5,2)_\text{o}$ at 1.5 K (red circles) and 40 K (blue circles). Right: Difference between the raw scans (black circles) and fit to gaussian profile (solid black line).}
	\label{fig:YTO_00P52}
	\end{figure}

\clearpage

	\subsection{$x = 0.10$}

\begin{figure*}[!htb]
	\includegraphics[width=0.6\textwidth]{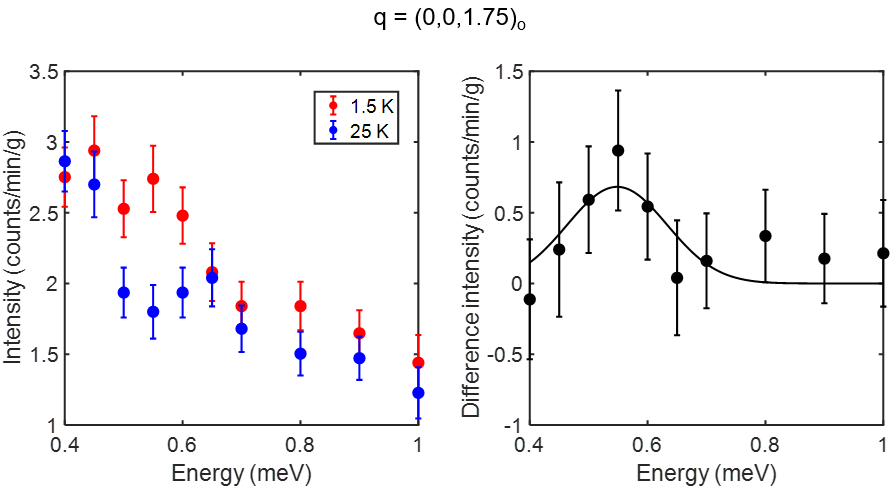}
	\caption{Left: Raw energy scans for $x = 0.10$ and $(0,0,1.75)_\text{o}$ at 1.5 K (red circles) and 25 K (blue circles). Right: Difference between the raw scans (black circles) and fit to gaussian profile (solid black line).}
	\label{fig:YLa10_001}
\end{figure*}

\begin{figure*}[!htb]
	\includegraphics[width=0.6\textwidth]{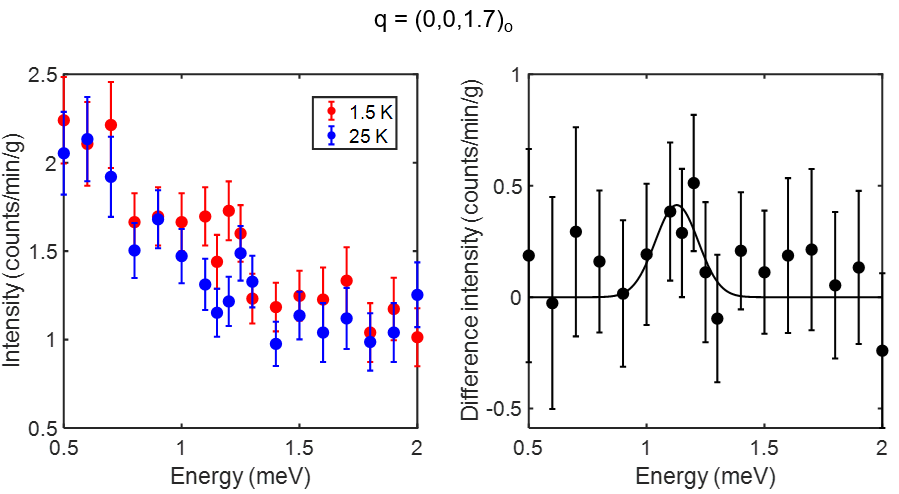}
	\caption{Left: Raw energy scans for $x = 0.10$ and $(0,0,1.7)_\text{o}$ at 1.5 K (red circles) and 25 K (blue circles). Right: Difference between the raw scans (black circles) and fit to gaussian profile (solid black line).}
	\label{fig:YLa10_001}
\end{figure*}

\begin{figure*}[!htb]
	\includegraphics[width=0.6\textwidth]{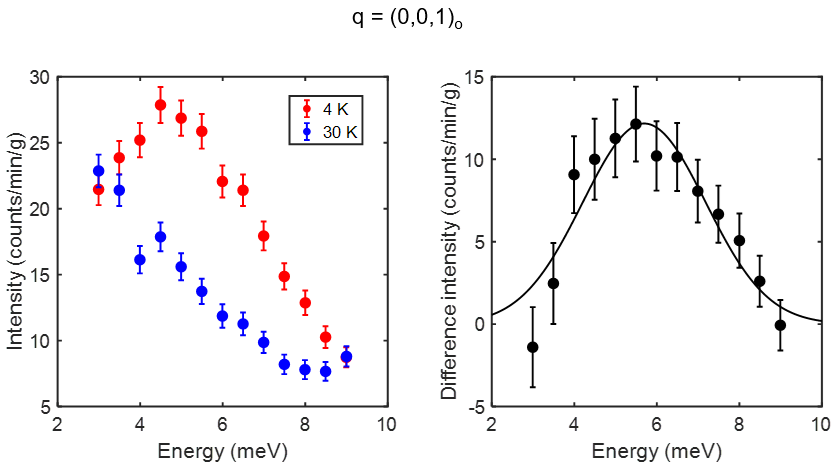}
	\caption{Left: Raw energy scans for $x = 0.10$ and $(0,0,1)_\text{o}$ at 4 K (red circles) and 30 K (blue circles). Right: Difference between the raw scans (black circles) and fit to gaussian profile (solid black line).}
	\label{fig:YLa10_001}
\end{figure*}

\begin{figure}[!htb]
	\includegraphics[width=0.6\textwidth]{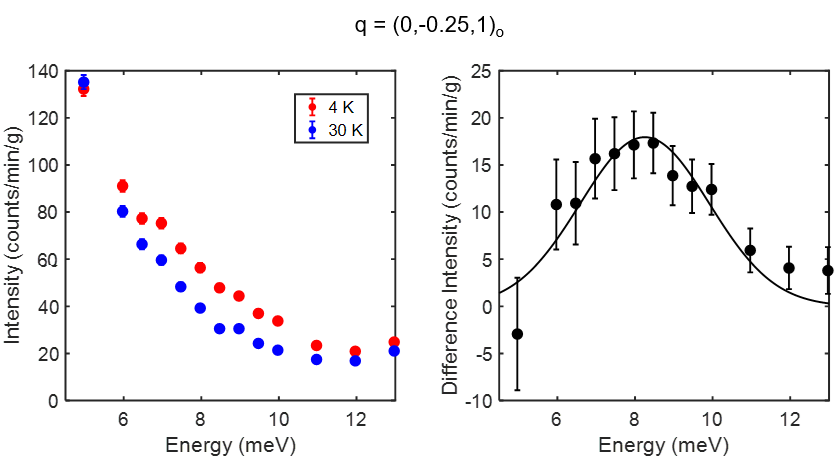}
	\caption{Left: Raw energy scans for $x = 0.10$ and $(0,-0.25,1)_\text{o}$ at 4 K (red circles) and 30 K (blue circles). Right: Difference between the raw scans (black circles) and fit to gaussian profile (solid black line).}
	\label{fig:YLa10_00P251}
\end{figure}

\begin{figure}[!htb]
	\includegraphics[width=0.6\textwidth]{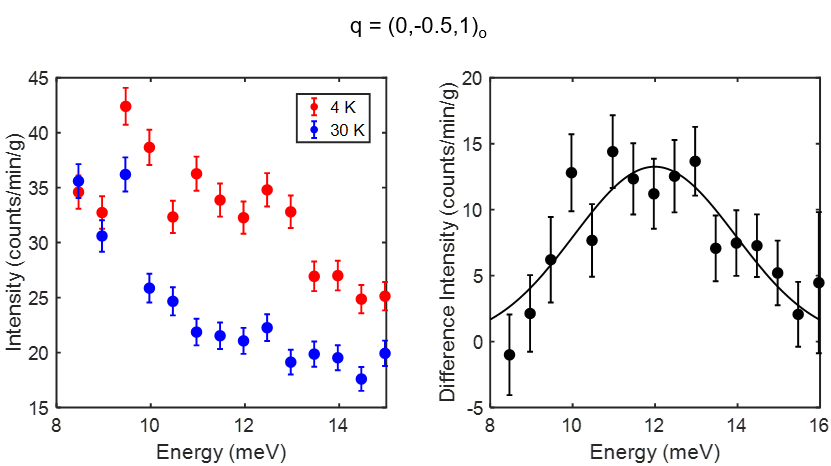}
	\caption{Left: Raw energy scans for $x = 0.10$ and $(0,-0.5,1)_\text{o}$ at 4 K (red circles) and 30 K (blue circles). Right: Difference between the raw scans (black circles) and fit to gaussian profile (solid black line).}
	\label{fig:YLa10_00P51}
\end{figure}

\begin{figure}[!htb]
	\includegraphics[width=0.6\textwidth]{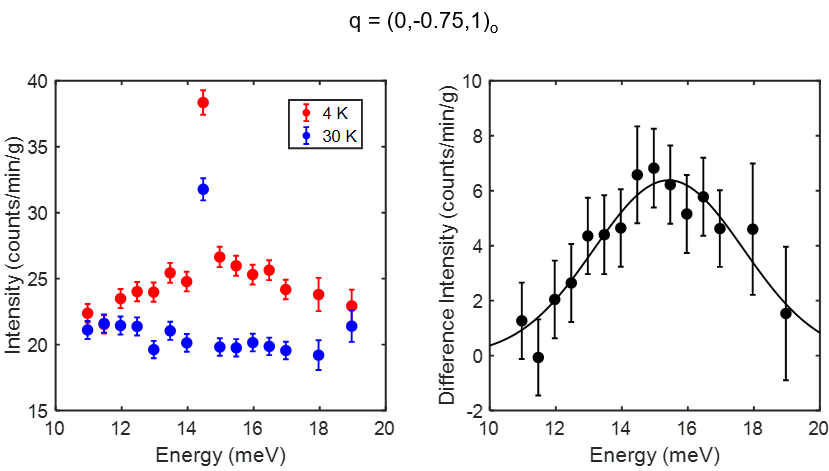}
	\caption{Left: Raw energy scans for $x = 0.10$ and $(0,-0.75,1)_\text{o}$ at 4 K (red circles) and 30 K (blue circles). Right: Difference between the raw scans (black circles) and fit to gaussian profile (solid black line).}
	\label{fig:YLa10_00P751}
\end{figure}

\begin{figure}[!htb]
	\includegraphics[width=0.6\textwidth]{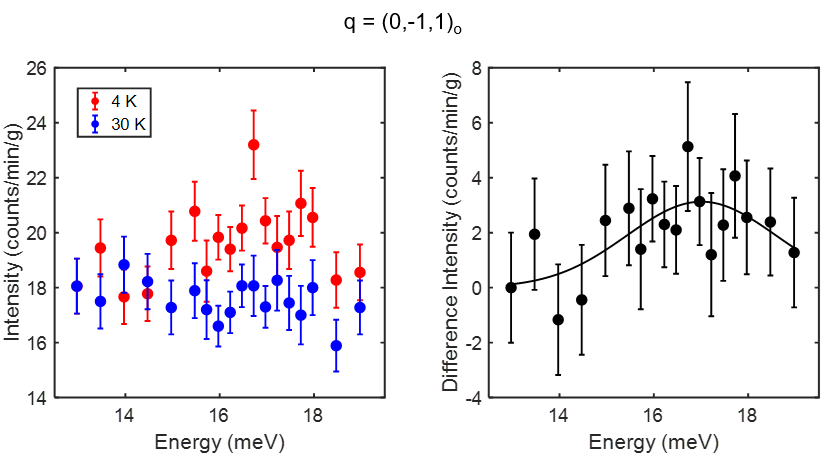}
	\caption{Left: Raw energy scans for $x = 0.10$ and $(0,-1,1)_\text{o}$ at 4 K (red circles) and 30 K (blue circles). Right: Difference between the raw scans (black circles) and fit to gaussian profile (solid black line).}
	\label{fig:YLa10_011}
\end{figure}

\begin{figure}[!htb]
	\includegraphics[width=0.6\textwidth]{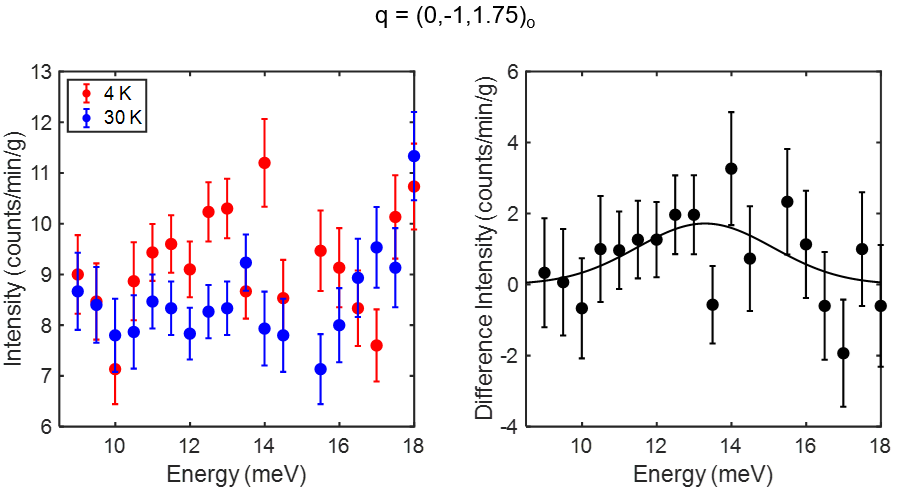}
	\caption{Left: Raw energy scans for $x = 0.10$ and $(0,-1,1.75)_\text{o}$ at 4 K (red circles) and 30 K (blue circles). Right: Difference between the raw scans (black circles) and fit to gaussian profile (solid black line).}
	\label{fig:YLa10_011P75}
\end{figure}

\begin{figure}[!htb]
	\includegraphics[width=0.6\textwidth]{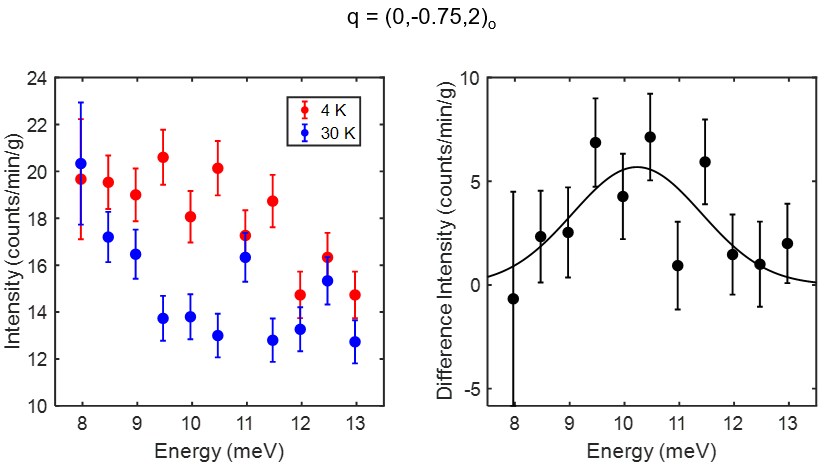}
	\caption{Left: Raw energy scans for $x = 0.10$ and $(0,-0.75,2)_\text{o}$ at 4 K (red circles) and 30 K (blue circles). Right: Difference between the raw scans (black circles) and fit to gaussian profile (solid black line).}
	\label{fig:YLa10_00P752}
\end{figure}

\begin{figure}[!htb]
	\includegraphics[width=0.6\textwidth]{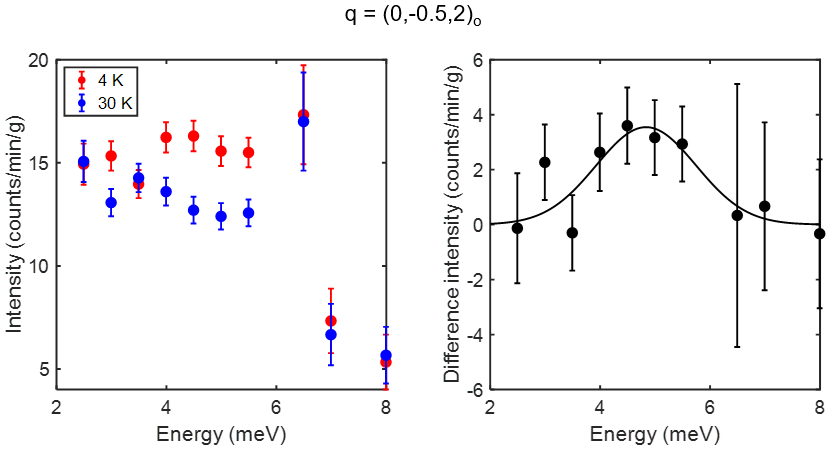}
	\caption{Left: Raw energy scans for $x = 0.10$ and $(0,-0.5,2)_\text{o}$ at 4 K (red circles) and 30 K (blue circles). Right: Difference between the raw scans (black circles) and fit to gaussian profile (solid black line).}
	\label{fig:YLa10_00P52}
\end{figure}

\begin{figure}[!htb]
	\includegraphics[width=0.6\textwidth]{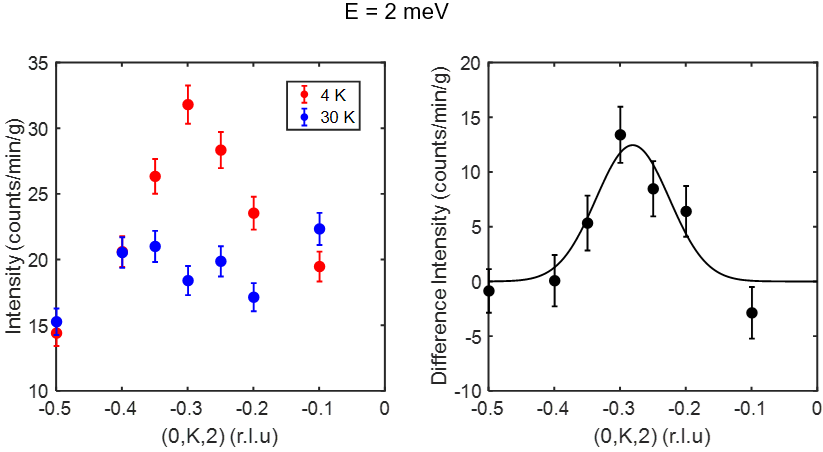}
	\caption{Left: Raw momentum scans for $x = 0.10$ and 2 meV at 4 K (red circles) and 30 K (blue circles). Right: Difference between the raw scans (black circles) and fit to gaussian profile (solid black line).}
	\label{fig:YLa10_0K2_2meV}
\end{figure}

\begin{figure}[!htb]
	\includegraphics[width=0.6\textwidth]{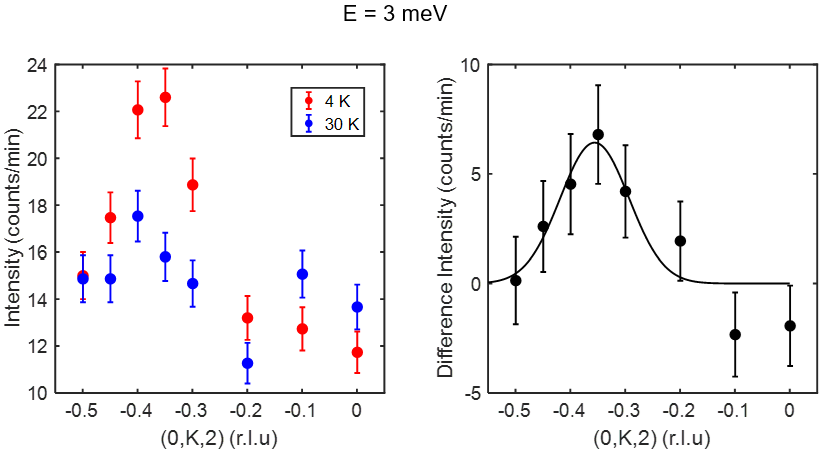}
	\caption{Left: Raw momentum scans for $x = 0.10$ and 3 meV at 4 K (red circles) and 30 K (blue circles). Right: Difference between the raw scans (black circles) and fit to gaussian profile (solid black line).}
	\label{fig:YLa10_0K2_2meV}
\end{figure}

\begin{figure}[!htb]
	\includegraphics[width=0.6\textwidth]{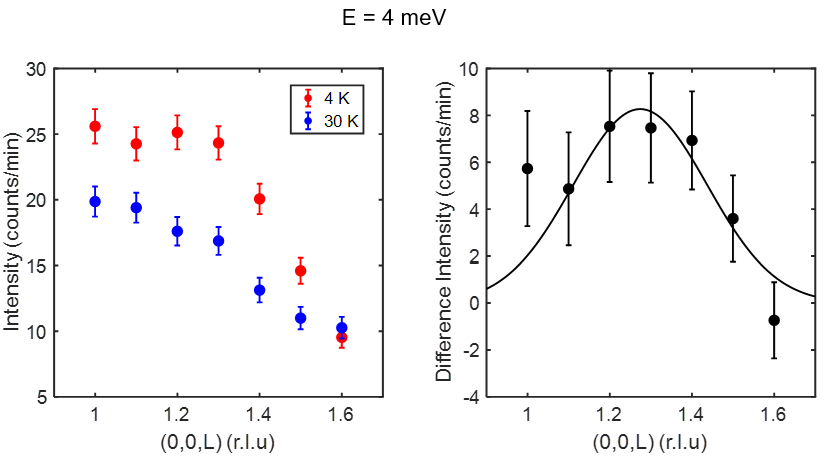}
	\caption{Left: Raw momentum scans for $x = 0.10$ and 4 meV at 4 K (red circles) and 30 K (blue circles). Right: Difference between the raw scans (black circles) and fit to gaussian profile (solid black line).}
	\label{fig:YLa10_00L_4meV}
\end{figure}

\clearpage

	\subsection{$x = 0.20$}

\begin{figure*}[!htb]
	\includegraphics[width=0.6\textwidth]{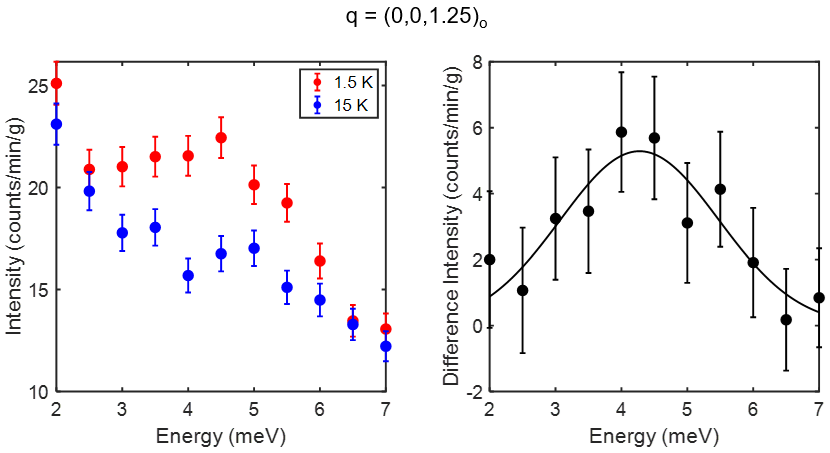}
	\caption{Left: Raw energy scans for $x = 0.20$ and $(0,0,1.25)_\text{o}$ at 1.5 K (red circles) and 15 K (blue circles). Right: Difference between the raw scans (black circles) and fit to gaussian profile (solid black line).}
	\label{fig:YLa20_001P25}
\end{figure*}

\begin{figure}[!htb]
	\includegraphics[width=0.6\textwidth]{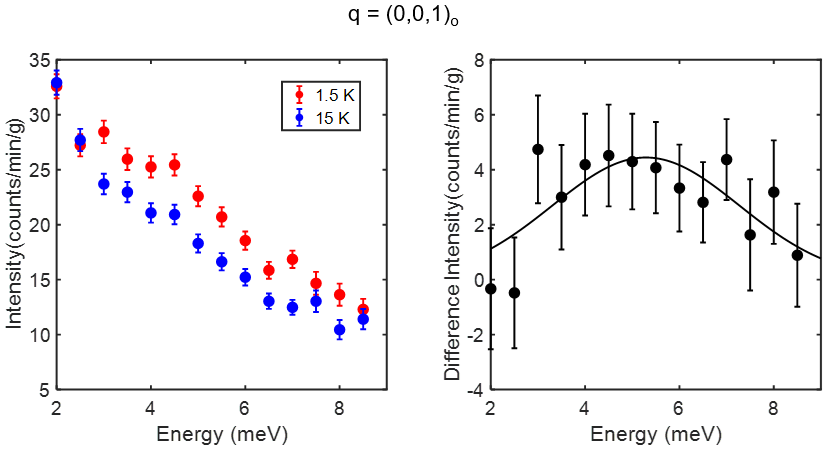}
	\caption{Left: Raw energy scans for $x = 0.20$ and $(0,0,1)_\text{o}$ at 1.5 K (red circles) and 15 K (blue circles). Right: Difference between the raw scans (black circles) and fit to gaussian profile (solid black line).}
	\label{fig:YLa20_001}
\end{figure}

\begin{figure}[!htb]
	\includegraphics[width=0.6\textwidth]{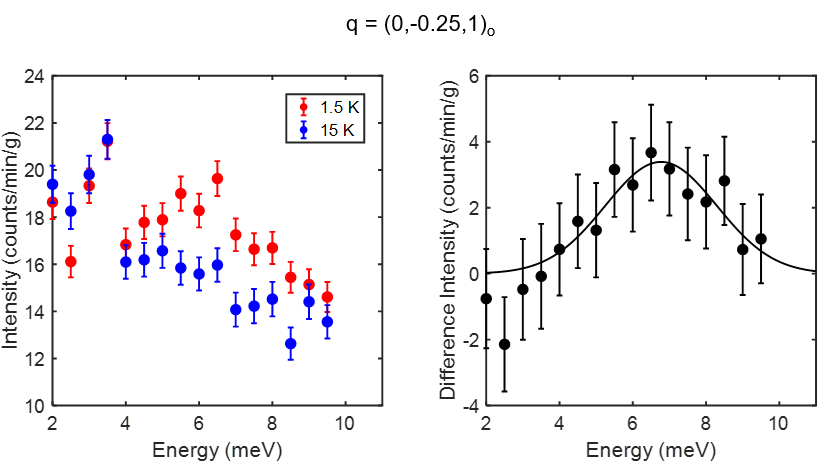}
	\caption{Left: Raw energy scans for $x = 0.20$ and $(0,-0.25,1)_\text{o}$ at 1.5 K (red circles) and 15 K (blue circles). Right: Difference between the raw scans (black circles) and fit to gaussian profile (solid black line).}
	\label{fig:YLa20_00P251}
\end{figure}

\begin{figure}[!htb]
	\includegraphics[width=0.6\textwidth]{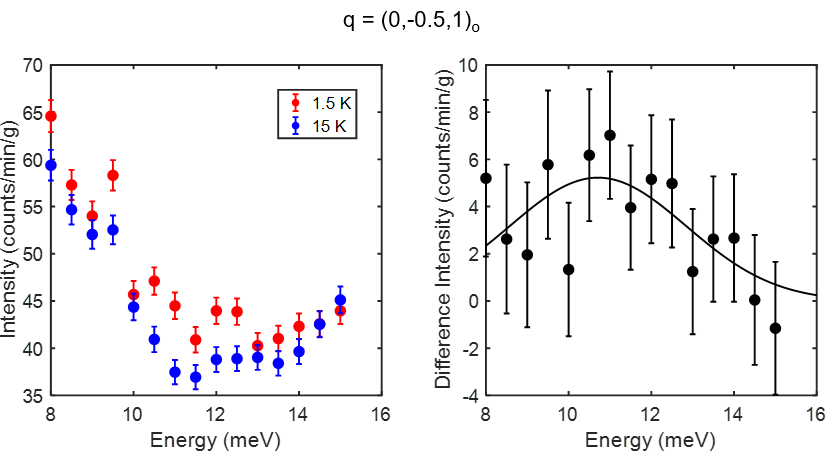}
	\caption{Left: Raw energy scans for $x = 0.20$ and $(0,-0.5,1)_\text{o}$ at 1.5 K (red circles) and 15 K (blue circles). Right: Difference between the raw scans (black circles) and fit to gaussian profile (solid black line).}
	\label{fig:YLa20_00P51}
\end{figure}

\begin{figure}[!htb]
	\includegraphics[width=0.6\textwidth]{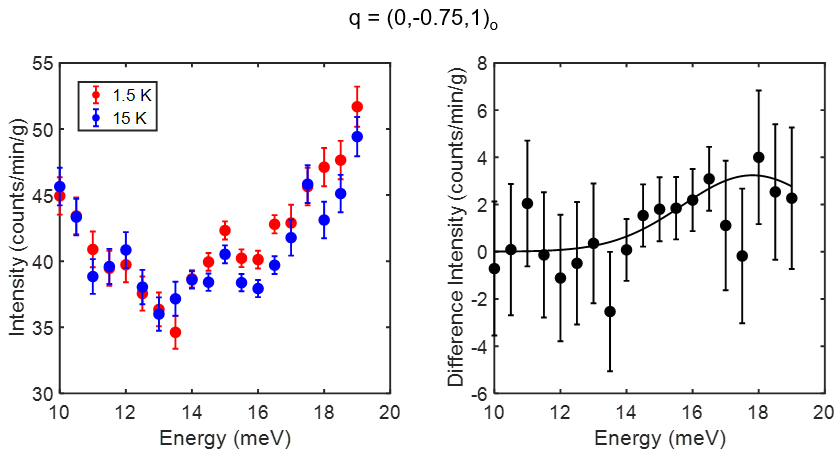}
	\caption{Left: Raw energy scans for $x = 0.20$ and $(0,-0.75,1)_\text{o}$ at 1.5 K (red circles) and 15 K (blue circles). Right: Difference between the raw scans (black circles) and fit to gaussian profile (solid black line).}
	\label{fig:YLa20_00P751}
\end{figure}

\begin{figure}[!htb]
	\includegraphics[width=0.6\textwidth]{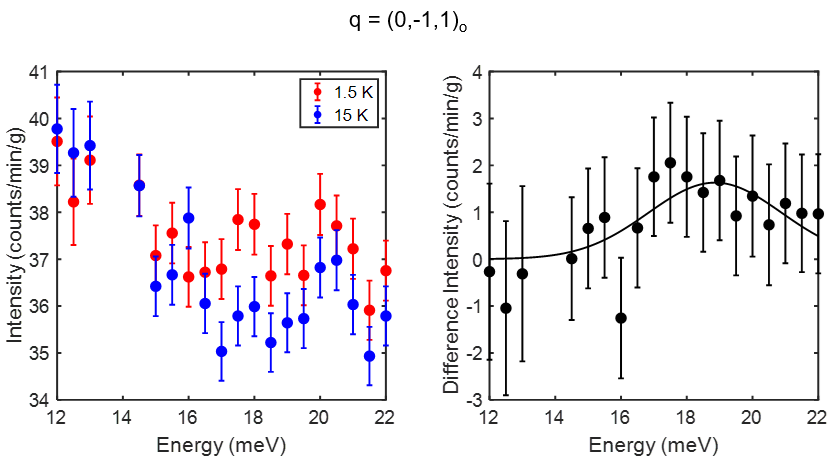}
	\caption{Left: Raw energy scans for $x = 0.20$ and $(0,-1,1)_\text{o}$ at 1.5 K (red circles) and 15 K (blue circles). Right: Difference between the raw scans (black circles) and fit to gaussian profile (solid black line).}
	\label{fig:YLa20_011}
\end{figure}

\begin{figure}[!htb]
	\includegraphics[width=0.6\textwidth]{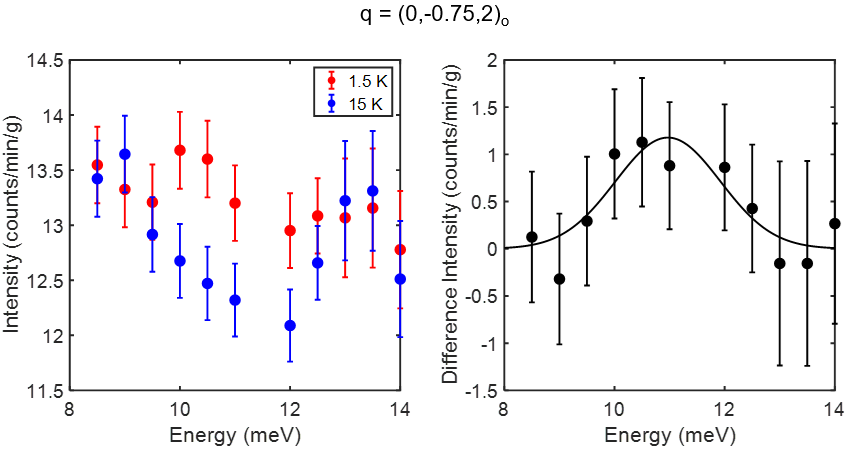}
	\caption{Left: Raw energy scans for $x = 0.20$ and $(0,-0.75,2)_\text{o}$ at 1.5 K (red circles) and 15 K (blue circles). Right: Difference between the raw scans (black circles) and fit to gaussian profile (solid black line).}
	\label{fig:YLa20_00P752}
\end{figure}

\begin{figure}[!htb]
	\includegraphics[width=0.6\textwidth]{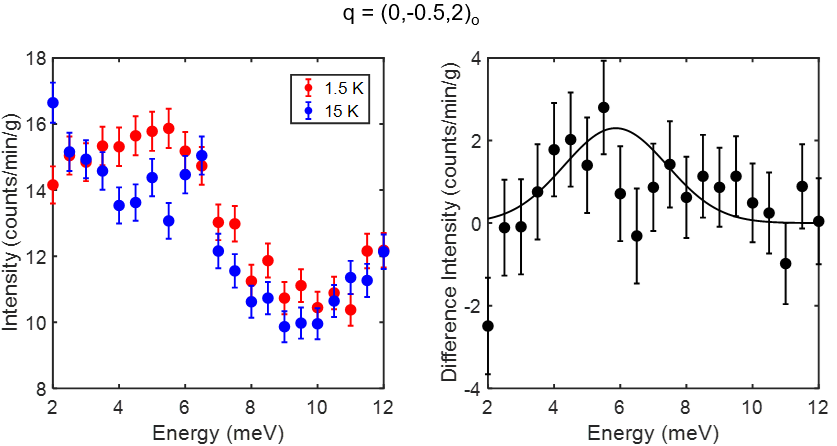}
	\caption{Left: Raw energy scans for $x = 0.20$ and $(0,-0.5,2)_\text{o}$ at 1.5 K (red circles) and 15 K (blue circles). Right: Difference between the raw scans (black circles) and fit to gaussian profile (solid black line).}
	\label{fig:YLa20_00P52}
\end{figure}

\begin{figure}[!htb]
	\includegraphics[width=0.6\textwidth]{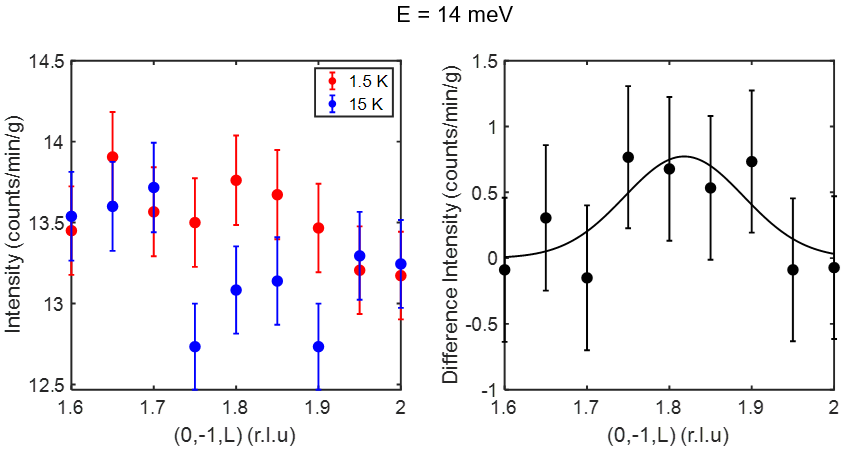}
	\caption{Left: Raw momentum scans for $x = 0.20$ and 14 meV at 1.5 K (red circles) and 15 K (blue circles). Right: Difference between the raw scans (black circles) and fit to gaussian profile (solid black line).}
	\label{fig:YLa20_0m1L_14meV}
\end{figure}

\clearpage

\section{Search for local antiferromagnetic spin waves for \lowercase{x} = 0.30}
\label{afmswsearch}
In prior theoretical work on FM systems with admixed AFM interactions, it was suggested that local AFM spin waves would form \cite{Ginzburg1979}. In this Section, we present attempts to find such local modes experimentally at and near the FM zone boundary in a $\sim$1.5 g $x = 0.30$ sample consisting of 11 comounted single crystals. Raw energy scans covering the low-energy region, taken at low temperature (1.5 K, below $T_C\sim~7$~K) and high temperature (15 K, above $T_C\sim~7$~K) are displayed. No clear spin wave signal is discernible from the scans at low energies, unlike what would be expected in the presence of local AFM spin waves.

\begin{figure}[!htb]
	\includegraphics[width=0.6\textwidth]{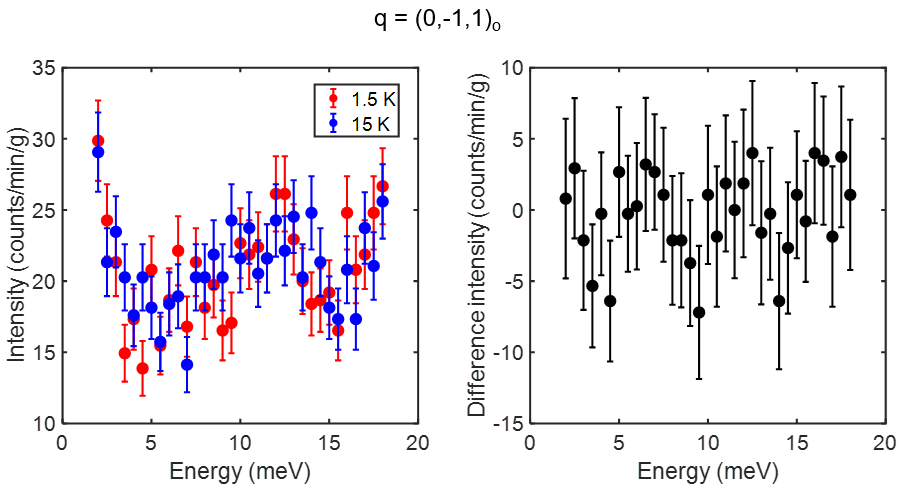}
	\caption{Left: Raw energy scans for $x = 0.30$ and $(0,-1,1)_\text{o}$ (FM zone boundary) at 1.5 K (red circles) and 15 K (blue circles). Right: Difference between the raw scans (black circles).}
	\label{fig:YLa30_011}
\end{figure}

\begin{figure}[!htb]
	\includegraphics[width=0.6\textwidth]{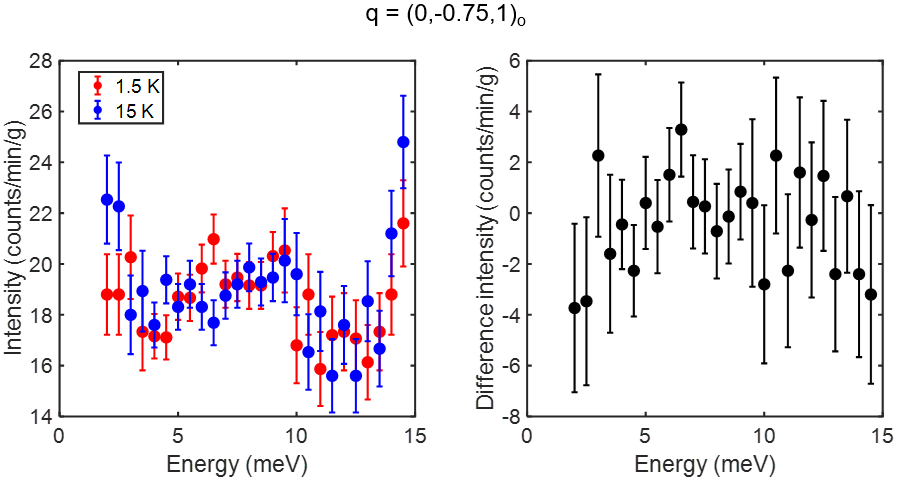}
	\caption{Left: Raw energy scans for $x = 0.30$ and $(0,-0.75,1)_\text{o}$ (near the FM zone boundary) at 1.5 K (red circles) and 15 K (blue circles). Right: Difference between the raw scans (black circles).}
	\label{fig:YLa30_00P751}
\end{figure}

\begin{figure}[!htb]
	\includegraphics[width=0.6\textwidth]{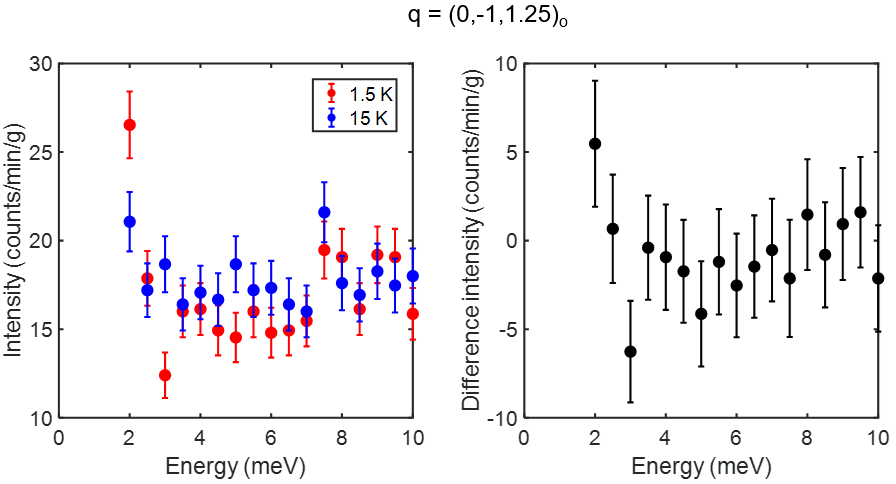}
	\caption{Left: Raw energy scans for $x = 0.30$ and $(0,-1,1.25)_\text{o}$ (near the FM zone boundary) at 1.5 K (red circles) and 15 K (blue circles). Right: Difference between the raw scans (black circles).}
	\label{fig:YLa30_011P25}
\end{figure}

\clearpage

\section{Comparison of spin-wave spectra for YT\lowercase{i}O$_3$ obtained from deconvolution with the instrument resolution and from simple Gaussian fits}
\label{ytodeconv}

	\begin{figure*}[!htb]
		\includegraphics[width=0.55\textwidth]{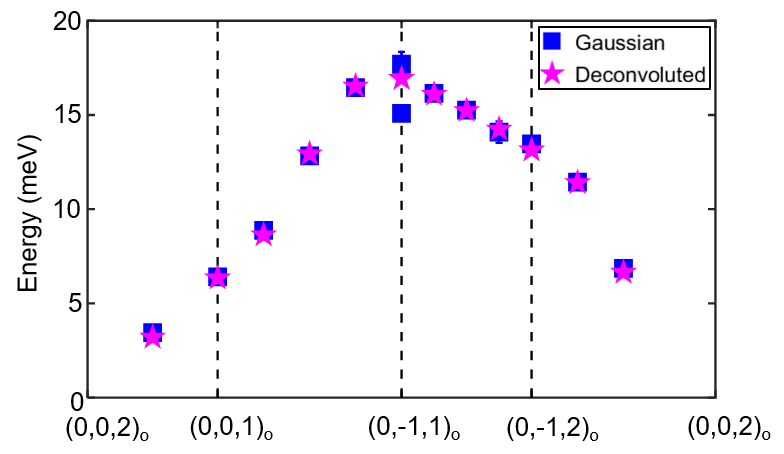}
		\caption{Comparison of spin-wave spectra for YTiO$_3$ obtained from a deconvolution of the spin-wave peaks with the instrument resolution function and from simple gaussian fits. The results are in excellent agreement.}
		\label{fig:deconvSW}
	\end{figure*}
	
\section{DFT Calculations}
\label{dftcalc}

All DFT calculations were performed using the projector augmented wave (PAW) formalism as implemented in the Vienna ab initio Simulation Package (VASP) \cite{Kresse1993,Kresse1996CMS,Kresse1996PRB}, employing a $2\times 2\times 2$ pseudocubic cell,  plane wave cutoff energy of 550 eV, and $4\times4\times4$ Monkhorst-Pack $k$-point grid. We use the PBEsol exchange correlation functional \cite{Perdew2008}, and employ the LSDA$+U$ scheme with $U=4$ eV set on the Ti $3d$ orbitals \cite{Dudarev1998}. 
	
\section{Monte Carlo simulation details}
\label{montecarlo}

A classical spin model on a simple cubic lattice with $L^3$ sites is considered:
\begin{equation}
	\mathcal{H} = \sum_{\langle ij\rangle} J_{ij} \bm{S}_i \cdot \bm{S}_j,
\end{equation}
where $\langle ij \rangle$ denotes the nearest neighbor bonds. As described in the main text, La doping introduces AFM bonds randomly distributed in the system. Since the N\'eel temperature $T_N$ in LaTiO$_3$ ($\sim 150$ K) is about 5 times the Curie temperature $T_C$ in YTiO$_3$ ($\sim 30$ K), we approximated $J_{ij}$ from a binary distribution: $J_{ij}=-J$ (FM) with probability $1-p$, and $J_{ij}=5J$ (AFM) with probability $p$. The spins are normalized with $|\bm{S}_i|=1$.

To compute the finite-temperature properties of the disorder model, we used the classical Monte Carlo (MC) calculation with parallel tempering technique~\cite{HukushimaK1996a}, combined with heat-bath~\cite{MiyatakeY1986} and over-relaxation updates~\cite{CreutzM1987,BrownFR1987}. Our unit MC step per spin includes one heat-bath sweep followed by $L$ over-relaxation sweeps. The inverse temperatures \{$\beta_i$\} ($\beta_1 < \beta_2 < \cdots$) were chosen such that the energy histograms of neighboring temperatures have sizeable overlap. After the $n$'th MC step ($n=1,2,3\ldots$), replica exchange trial was performed between neighboring pairs of temperatures, except that the spin configurations with $\beta_i$ satisfying $\text{mod}(i+n,3)=0$ did not participate in the exchange trial.

For $L=16$, we used $10^6$ MC steps per spin to equilibrate the system, and another $10^6$ MC steps per spin for measurement. The minimal temperature in the parallel tempering was chosen as $T_\text{min}/J=0.1$, and we typically used about 50 temperatures to make sure the histograms between neighboring temperatures have sizeable overlap. Finally, 50 disorder realizations were used for averaging the dynamic spin structure factor. 

\begin{figure}
	\includegraphics[width=0.5\columnwidth]{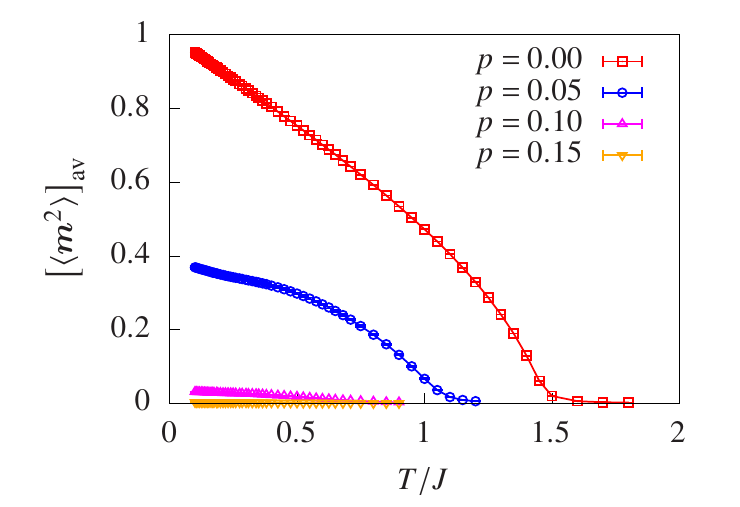}
	\caption{The squared magnetization $\left[\langle \bm{m}^2 \rangle\right]_\text{av}$ as a function of temperature, computed for $L=16$ and $p=\{0,\, 0.05,\, 0.1, \, 0.15 \}$. }
	\label{fig:magnetization_MC}
\end{figure}

Figure.~\ref{fig:magnetization_MC} shows the squared magnetization $\left[\langle \bm{m}^2 \rangle\right]_\text{av}$ as a function of temperature, which indicates that $p=0.1$ is close to the experimental data for $x = 0.2$. We note that such a comparison may not be quantitatively accurate since the model we use is a simplified one. 

In the presence of finite disorder, it is convenient to compute the spin wave excitations by the equation-of-motion method. After obtaining the equiliberated spin configurations at the lowest temperature $T/J=0.1$, we evolve the spins by the Landau-Lifshitz (LL) equation:
\begin{equation}
	\frac{\mathrm{d}\bm{S}_i}{\mathrm{d}t}= -\bm{S}_i \times \frac{\mathrm{d}E}{\mathrm{d}\bm{S}_i},
\end{equation}
where $E$ is the internal energy of the system.

Then we Fourier transformed this series of spin configurations to obtain the dynamical spin structure factor:
\begin{equation}
	\mathcal{S}(\bm{k},\omega) = \frac{\omega}{T} \left[ \langle \bm{S}_{\bm{k}}(\omega) \cdot \bm{S}_{-\bm{k}}(-\omega) \rangle \right]_\text{av},
\end{equation}
where the Fourier components are
\begin{equation}
	\bm{S}_{\bm{k}}(\omega) \equiv \frac{1}{\sqrt{\mathcal{T}}}\int_0^\mathcal{T} \mathrm{d}t e^{\iu \omega t} \frac{1}{L^{3/2}}\sum_i e^{-\iu \bm{k} \cdot \bm{r}_i } \bm{S}_i (t).
\end{equation}

Here, we integrated the LL equation in the 4th order Adams predictor-corrector scheme, with a total of $2\times 10^4$ steps of duration $\Delta t = 0.01 J^{-1}$ ($\mathcal{T} = 200 J^{-1}$), and used 50 realizations for the disorder average. The computed $\mathcal{S}(\bm{k},\omega)$ are presented in Fig.~\ref{fig:disorder} of the main text. Note that $\omega$ in $\mathcal{S}(\bm{k},\omega)$ should be rescaled by $S$ when we choose a different normalization of spin length.

\end{document}